\documentstyle[11pt,aasms4,rotating]{article}

\def\simlt{\lower.5ex\hbox{$\; \buildrel < \over \sim \;$}}
\def\simgt{\lower.5ex\hbox{$\; \buildrel > \over \sim \;$}}

\def\gcm3{{\rm\,g\,cm^{-3}}}
\def\ncm3{{\rm\,cm^{-3}}}

\def\>{$>$}
\def\<{$<$}

\lefthead{Melia et al.}
\righthead{}

\begin{document}
\centerline{Submitted to the Editor of the Astrophysical Journal Letters}
\vskip 0.5in
\title{\bf Measuring the Black Hole Spin in Sgr A*}

\author{Fulvio Melia$^{1,2,3}$, Benjamin C. Bromley$^{4}$, Siming Liu\altaffilmark{2},
and Christopher, K. Walker\altaffilmark{3}}

\affil{$^2$Physics Department, The University of Arizona, Tucson, AZ 85721}
\affil{$^3$Steward Observatory, The University of Arizona, Tucson, AZ 85721}
\affil{$^4$Department of Physics, University of Utah, 201 JFB, Salt Lake City, 
UT 84112}

% Notice that each of these authors has alternate affiliations, which
% are identified by the \altaffilmark after each name.  The actual alternate
% affiliation information is typeset in footnotes at the bottom of the
% first page, and the text itself is specified in \altaffiltext commands.
% There is a separate \altaffiltext for each alternate affiliation
% indicated above.

\altaffiltext{1}{Sir Thomas Lyle Fellow and Miegunyah Fellow.}

% The abstract environment prints out the receipt and acceptance dates
% if they are relevant for the journal style.  For the aasms style, they
% will print out as horizontal rules for the editorial staff to type
% on, so long as the author does not include \received and \accepted
% commands.  This should not be done, since \received and \accepted dates
% are not known to the author.

\begin{abstract}
The polarized mm/sub-mm radiation from Sgr A* is apparently produced by a 
Keplerian structure whose peak emission occurs within several
Schwarzschild radii ($r_S\equiv 2GM/c^2$) of the black hole.  The
{\it Chandra} X-ray counterpart, if confirmed, is presumably the
self-Comptonized component from this region.  In this paper, we suggest
that sub-mm timing observations could yield a signal
corresponding to the period $P_0$ of the marginally stable orbit,
and therefore point directly to the black hole's spin $a$.  Sgr A*'s
mass is now known to be $(2.6\pm 0.2)\times 10^6\,M_\odot$ (an unusually
accurate value for supermassive black hole candidates), 
for which $2.7\;\hbox{\rm min}<P_0<
36$ min, depending on the value of $a$ and whether the Keplerian
flow is prograde or retrograde.  A Schwarzschild black hole ($a=0$)
should have $P_0\approx 20$ min. The identification of the 
orbital frequency with the innermost stable circular orbit is made
feasible by the transition from optically thick to thin emission at
sub-mm wavelengths.  With stratification in the emitter, the peak of 
the sub-mm bump in Sgr A*'s spectrum is thus produced at the smallest 
radius.  We caution, however, that theoretical uncertainties in the
structure of the emission region may still produce some ambiguity
in the timing signal. Given that Sgr A*'s flux at $\nu\sim 1$ mm is several 
Jy, these periods should lie within the temporal-resolving capability of 
sub-mm telescopes using bolometric detectors. A determination of $P_0$ 
should provide not only a value of $a$, but it should also define the 
angular momentum vector of the orbiting gas in relation to the
black hole's spin axis.  By analogy with low-mass X-ray binaries and
Galactic black hole candidates, Sgr A* may also display quasi-periodic
oscillations, which can reveal additional features in the geometry of
the accreting gas. In addition, since the X-ray flux detected by {\it Chandra}
appears to be the self-Comptonized mm to sub-mm component, these
temporal fluctuations may also be evident in the X-ray signal.
\end{abstract}

% The different journals have different requirements for keywords.  The
% keywords.apj file, found on aas.org in the pubs/aastex-misc directory, 
% contains a list of keywords used with the ApJ and Letters.  These are 
% usually assigned by the editor, but authors may include them in their 
% manuscripts if they wish. 

\keywords{accretion---black hole physics---Galaxy: 
center---hydrodynamics---magnetic 
fields: dynamo---radiation mechanisms: nonthermal}

% That's it for the front matter.  On to the main body of the paper.
% We'll only put in tutorial remarks at the beginning of each section
% so you can see entire sections together.

% In the first two sections, you should notice the use of the LaTeX \cite
% command to identify citations.  The citations are tied to the
% reference list via symbolic KEYs.  We have chosen the first three
% characters of the first author's name plus the last two numeral of the
% year of publication.  The corresponding reference has a \bibitem
% command in the reference list below.
%
% Please see the AASTeX manual for a more complete discussion on how to make
% \cite-\bibitem work for you.   

\section{Introduction}
It is now thought that the dynamical center of the Galaxy coincides with
Sgr A* (Eckart et al. 1995; Menten et al. 1997; Ghez et al. 1998), a compact
nonthermal radio source no bigger than $\sim 1$ AU (Krichbaum et al. 1993;
Backer et al. 1993; Krichbaum et al. 1998; Lo et al. 1993).  With a concentration
of dark matter ($2.6\pm0.2\times 10^6\;M_\odot$) within 0.015 pc of its centroid
(Genzel et al. 1996; Ghez et al.  1998),  Sgr A* anchors the stars and gas 
locked in its vicinity, and provides possibly the most compelling evidence for 
the existence of supermassive black holes.  Its time-averaged spectrum is
roughly a power-law, $S_{\nu}\propto\nu^a$, with $a\sim 0.19-0.34$ from 
cm to mm wavelengths.  In the sub-millimeter (sub-mm) region, a spectral excess
(or ``bump'') has emerged (Zylka et al. 1992; Zylka et al.  1995), which
now appears to be well established (Falcke, et al. 1998). This is
interesting in view of the fact that it may be a signature of activity
close to the black hole's event horizon, since the highest frequencies 
appear to correspond to the smallest spatial scales (e.g., Melia, Jokipii 
\& Narayanan; Melia 1992, 1994; Narayan et al. 1995; Falcke et al. 1998;
Coker \& Melia 2000).

The detection of Sgr A* at X-ray wavelengths (Baganoff et al. 2001), 
together with the radio polarization 
measurements reported by Bower et al. (1999) and Aitken et al. (2000), offer
the best constraints yet for determining the nature of the emitting gas,
and understanding the environment within several Schwarzschild radii 
($r_S\equiv 2GM/c^2=7.7\times 10^{11}$ cm) of the black hole. This is because
the behavior of Sgr A* is dictated by the manner with which plasma accretes 
onto it from the nearby medium. Through a series of large scale 3D
hydrodynamic simulations (e.g., Coker \& Melia 1997), it has become
apparent that the specific angular momentum $\lambda$ (in units of $cr_S$) 
accreted with the gas is variable (on a time scale of decades) and
rather small, averaging $<40$. This is not sufficient for the plasma to
form a large disk (such as that required by ADAF disk models; Narayan et 
al. 1995), but it is expected to circularize the inflow at a radius
$r_{circ}=2\lambda^2\,r_S$ before spiraling in through the event horizon.

The existence of a Keplerian structure at small radii should still
be viewed as an assumption.  The reason for this is that the hydrodynamic 
simulations conducted thus far have necessarily employed an inner boundary 
with a radius ($r_{in}>1,000 r_S$) much larger than $r_{circ}$. However, a 
small Keplerian structure is implied by the detailed modeling of Sgr A*'s 
spectrum (from radio to X-rays), which has demonstrated that the sub-mm 
``excess'' of emission seen in its spectrum may be associated with radiation 
produced within this region of circularized flow (Melia, Liu \& Coker 2000, 
2001).  The sub-mm emission probably 
represents a transition from optically thick to optically thin emission 
(Melia 1992, 1994), so that radiation below the peak of the sub-mm bump 
originates predominantly in front of and behind the black hole, whereas
radiation above the peak is produced by (and Doppler boosted within) the 
orbiting gas to the side of the central object.  The implied degree of 
polarization (about $10\%$) and a rotation in the position angle (by almost 
$90^\circ$) across the bump are consistent with what has been seen in this 
source (Aitken et al. 2000).  The {\it Chandra} observations are crucial for 
establishing the viability of this picture because the hot Maxwellian particles 
producing the cm to sub-mm radiation via thermal synchrotron emission can
also self-Comptonize the soft radiation to produce X-rays. 

These three pieces of circumstantial evidence (the mm/sub-mm bump, its polarization
characteristics and the self-Comptonized X-ray component) now suggest the 
intriguing possibility that we may be on the threshold of actually measuring 
the spin of the black hole in Sgr A*.  Since its mass is known to such high 
precision, features seen in the power spectrum for the mm/sub-mm emission can 
directly yield the factors defining the circum-black hole geometry.

\section{A Possible Periodicity in Sgr A*'s Millimeter/Sub-Millimeter Spectrum}
The structure of the circularized flow within $\sim 5-50\;r_S$ is developed 
fully in Melia, Liu \& Coker (2000).  Central to the modeling of the sub-mm 
``excess'' is the supposition that within the Keplerian flow, a 
magnetohydrodynamic dynamo produces an enhanced (though still sub-equipartition) 
magnetic field, dominated by its azimuthal component (Hawley, Gammie \& Balbus 
1996).  Briefly, Melia, Liu \& Coker (2000) infer the following physical state of the
gas toward small radii (i.e., within $\sim 5r_S$ or so).  The polarization data
(particularly those presented by Aitken et al. 2000) appear to restrict the
accretion rate to values no larger than about $10^{16}$ g s$^{-1}$, for otherwise
the medium would not become transparent at sub-mm wavelengths.  In order to account
for the sub-mm bump with thermal synchrotron emission, the temperature $T$ (assumed to
be the same for the electrons and the protons) should be relativistic.  In the
best fits, the profile in $T$ shows a steady rise from $\sim 10^{10}$ K at the
outer edge of the Keplerian flow to about $10^{11}$ within the last stable orbit.
Following the prescription for calculating the magnetic field and the viscosity 
(e.g., Hawley et al. 1996), Melia, Liu \& Coker (2000) determined the corresponding 
particle density $n$ and magnetic field intensity $B$ self-consistently. In the 
best fits to the data, they found that $2\times 10^6\;{\hbox{cm}}^{-3} <n<2\times 
10^7$ cm$^{-3}$, and $6\;\hbox{G}<B<17$ G.  This range for $B$ is consistent with 
the result of Beckert \& Duschl (1997) and Coker \& Melia (2000).  In the sub-mm 
region, the thermal synchrotron flux density produced by this gas peaks at 
$2.4\times 10^{11}$ Hz, and the flip frequency (where the position angle shifts by 
about $90^\circ$) is $2.8\times 10^{11}$ Hz. 

The sub-mm bump represents a transition from optically thick to thin emission.
Thus, since the dominant frequency of emission increases with decreasing radius, 
the peak of this bump is expected to be produced at the smallest stable radius.  
Above this frequency, the medium is transparent.  If this picture is
correct, general relativistic effects should significantly influence Sgr A*'s 
mm/sub-mm spectrum (Bromley, Melia \& Liu 2001), which would provide an
observational signature that can be tested with high-sensitivity single-dish
observations.  Chief among these effects is the expected amplification of 
nonaxisymmetric modulations in the disk or jet flux due to strong light 
bending and Doppler shift corrections within $\sim 4-5$ Schwarzschild radii of the
black hole, which might therefore produce a characteristic period, associated
with the innermost stable circular orbit.  Earlier
attempts to see a periodicity in Sgr A*'s spectrum were concerned with its
possible infrared emission (Hollywood et al. 1995), which we now know is very
weak given the very high temperatures and low density attained by the gas
near the event horizon.  However, since the X-rays appear to be produced
by self-Comptonization of the radio photons, it may be possible to see these
temporal variations in the X-ray signal as well.

More rapid disk fluctuations may also be present, which would stand out
in the power spectrum as peaks associated with quasi-periodic oscillations (QPO).
Reports of periodic or quasi-periodic oscillations (QPO) in Active
Galactic Nuclei have been rare, and in at least one case (NGC6814)
were later shown to be due to confusion with a nearby object 
(Madejski, et al. 1993).  But in the case of NGC5548 (Papadakis
\& Lawrence 1993), the QPOs with period $\sim 500$ s do appear to
be intrinsic to the source. 

X-ray observations of low-mass X-ray binaries (LMXBs) 
containing low-magnetic-field neutron stars have revealed 
two simulataneous quasi-periodic oscillation peaks in the 300-1300 Hz range 
some 300 Hz apart (van der Klis et al. 1996; Strohmayer et al. 1996).
The characteristics of these frequencies argue strongly in favor of a 
beat-frequency interpretation, in which the higher frequency ($\nu_{\rm orb}$) 
is associated with some preferred radius in the accretion disk, while the 
lower one is the beat frequency ($\nu_{\rm orb}-\nu_{\rm s}$) between
$\nu_{\rm orb}$ and the neutron star's spin frequency $\nu_{\rm s}$.
However, the fact that the separation between the two peaks (which is 
supposed to be equal to $\nu_{\rm s}$) is not exactly constant, has 
triggered other possible interpretations. The so-called relativistic 
precession model (Stella \& Vietri 1998) takes the upper frequency to be 
an orbital frequency, whereas the lower one is now the frequency of a
general-relativistic precession mode of a free particle orbit at that
radius.  Regardless, all models of the LMXBs QPO are based on the interpretation
that one of the kHz frequencies is an orbital frequency in the disk.

The idea of using kHz QPO to constrain neutron-star masses and radii,
and to test general relativity, traces its roots to the work of
Klu\'zniak \& Wagoner (1985), Paczy\'nski (1987), and Biehle \& 
Blandford (1993), among others.  Interestingly, the maximum kHz QPO frequencies
observed in each source are constrained to a narrow range.  Zhang
et al. (1997) proposed that this narrow distribution is caused by
the limit set by the innermost stable circular orbit.  Miller et al. (1998)
have suggested that when the inner edge of the accretion disk reaches this 
orbit, the QPO frequency should level off and remain constant even when other 
accretion parameters (such as the accretion rate) continue to change.  Some 
evidence for this has been seen in 4U 1820-30 (Zhang et al. 1998; Kaaret et al.
1999).  

The situation with the black hole candidates (BHCs) in binary systems is more 
complex and not as well understood.  Oscillations with frequencies
in the range $100-300$ Hz have been seen in several sources (including Cyg X-1),
and they tend to be constant.  This has motivated interpretations in which
these frequencies depend mostly on black-hole mass and angular momentum,
e.g., through the orbital motion at the innermost stable circular orbit
(Morgan et al. 1997), the Lense-Thirring precession at that radius
(Cui et al. 1998), and trapped-mode disk oscillations (Nowak et al. 1997).
The situation with Sgr A* is unique, and presumably better
suited for such studies, for several reasons, not the least of which is the 
fact that its mass is known very accurately.  In addition, whereas the X-ray
emitting gas in LMXBs and BHCc is presumably thick over a large range of
radii, the sub-mm bump in Sgr A* is produced at the smallest
radius (see above), which should confine the variability frequency to that
of the innermost stable circular orbit. We caution, however, that the
inferences we draw from the timing observations of Sgr A* may still be
somewhat ambiguous given the theoretical uncertainties regarding the
structure of the emission region in this source.  For example, if the range
of radii associated with the temporal variations is afterall larger than
what we are hypothesizing here, identification of the compact object's spin
would be subject to the same limitations as those described above for the 
LMXB. As we shall see below, it may nonetheless be possible not only to
infer Sgr A*'s spin, but also to identify whether the Keplerian flow is in 
prograde or retrograde rotation. 

\section{Identifying the Observed Variability}
The most rapid variability is expected to occur at the marginally
stable (circular) orbit, $r_{ms}(a)$ (see Bardeen, et al. 1972)
where $a$ is the black hole's angular momentum parameter.  This orbit
has a radius $r_{ms}(0)=3r_S$ for a Schwarzschild black hole, in which $a=0$. 
However, if the black hole is rotating, the location of the innermost
stable orbit (including the effects of frame dragging) is dictated
by the relative orientation of the spin angular momentum vector
of the accretor and that of the infalling material.  For illustrative
purposes, we will consider the situation for a maximally rotating
object, for which $a=r_G\equiv GM/c^2$.  In that case, the prograde ($+$) and
retrograde ($-$) marginally stable orbits have radii, respectively,
$r_{ms}^+(r_G)=r_G$ and $r_{ms}^-(r_G)=9r_G$.

Since for circular equatorial orbits in the Kerr metric, Kepler's
Third Law takes the form
\begin{equation}\label{eq:kepler}
\Omega={2\pi\over P}=\pm{cr_G^{1/2}\over r^{3/2}\pm a\,r_G^{1/2}}\;,
\end{equation}
the most rapid fluctuations associated with Sgr A* are expected to
have a period somewhere within the following extreme values:
$P_{min}(a=r_G,+)\approx 2.7\;\hbox{minutes}\;\left(M/
2.6\times 10^6\;M_\odot\right)$, and
$P_{min}(a=r_G,-)\approx 36.3\;\hbox{minutes}\;\left(M/
2.6\times 10^6\;M_\odot\right)$, with
$P_{min}(a=0)\approx 19.9\;\hbox{minutes}\;\left(M/
2.6\times 10^6\;M_\odot\right)$.
In other words, depending on its spin, Sgr A* is expected to
show variations on a time scale of $\sim 3-30$ minutes
at a spectral frequency $\nu\sim 2\times 10^{11}$ Hz.

As such, we should be able to distinguish this from other
(known) possible sources of variability, all of which have
a considerably longer time scale than this: stellar opacity
changes typically occur over months to years (e.g., long-period
variables);  a wind source fluctuation would occur over 1 day
(e.g., for IRS16NW, the Wolf-Rayet star closest to Sgr A*, the
time scale would be $R_{star}/v_{wind} \sim 18$ hours using the
parameters from Najarro, et al. 1997); stellar rotations,
sunspot activity, and binary motion all would last for days;
and the gravitational lensing of stars passing behind the
black hole would take weeks to years to produce a noticeable
variation (e.g., Alexander \& Sternberg 1998).  In addition,
the fluctuations associated with lensing and accretion phenomena
should be achromatic, in contrast to the other forms of variability 
discussed above, and should occur at the position of Sgr A* within 
the instrument's spatial resolution. 

\section{Sample Light Curves and Discussion}
To illustrate the possible mm and sub-mm variability in Sgr A*, we 
here adopt the specific model for the gas accretion near the event 
horizon described in \S\ 2, which is based on the magnetohydrodynamic 
dynamo picture developed by Melia, Liu \& Coker (2000, 2001).  We 
have chosen a Keplerian structure that runs from an inner boundary 
near $r_S$ to $5\,r_S$; the accretion flow is on circular orbits,
except within $r_{ms}(a)$, where the gas is assumed to be on freefall 
trajectories origination from just within the marginally stable orbit. 
The dynamo model then provides temperature, density and magnetic field 
strength in local reference frames which comove with the gas. From 
these we calculate the local synchrotron emissivity (e.g., Pacholczyk 
1970).

The accretion flow resides in a strong gravitational field so that
relativistic beaming, Doppler shifting and gravitational focusing all
modulate the flux measured well away from Sgr A*. Here we use a fully
relativistic ray tracing code (Bromley, Chen \& Miller 1997) to
determine how these effects influence the synchrotron emission as it
propagates from a comoving frame to a distant detector, idealized as
an array of pixels. We obtain the specific intensity $I$ at observed
frequency $\nu$ for each pixel by tracing photon trajectories back to
the disk and using the Lorentz invariant $I/\nu^3$. The frequency
of the photons and their emission angles in the comoving frame
come from projection of the photon 4-momentum onto a tetrad basis
tied to the emitter (e.g., Kojima 1991). The observed flux then follows
from integration over pixels in the detector array.

Flux variations will presumably result from localized perturbations
within the disk. For simplicity, we model such a perturbation by 
raising the temperature within a wedge-shaped patch of gas on the 
disk by 30\% while leaving all other model parameters unmodified.  
The wedge rotates as a pattern with angular velocity pegged to the 
Keplerian value at the innermost stable circular orbit, since most
of the emission at the peak frequency is produced within a narrow
range of radii there (see above).  We obtain the lightcurve from the 
patch by tracking position along its orbit and the time of flight for 
photons emitted within the patch.  

We consider three black hole spin values, $a=0$ and $a=\pm0.9$.
For the cases with non-zero angular momentum, the disk is assumed to
lie in the equatorial plane of the black hole. The wedge-shaped
perturbation is 45$^\circ$ in azimuthal extent,
and its inner edge is at or near the marginally stable orbit.
The figure caption indicates the exact radial boundaries in
each of the three systems.
Figure 1 illustrates the lightcurves obtained in this
way at $\nu=3\times 10^{11}$~Hz for a disk inclined at 60$^\circ$ 
relative to the plane of the sky.  Even the modest perturbations 
considered here can yield a detectable modulation in the lightcurve
(see below).  The weakest signal comes from the counterrotating disk, 
since in this case the perturbation is squeezed into a relatively small 
area between the outer edge of the disk and the marginally stable orbit. 
In all cases, the variability 
increases with frequency in mm and sub mm wavebands.

A key question concerns the feasibility of measuring the effects
we have described above.  The detection of the predicted flux 
variability in Sgr A* appears to be well within the capabilities of 
bolometric systems on existing mm/sub-mm telescopes. For the limiting 
case where $a = 0$, the predicted flux variation over the minimum
predicted period, $P_{min} = 19.9$~minutes, is $\sim 80$~mJy. 
On a $10$~m class submillimeter telescope, a typical bolometric
sensitivity at 0.87 mm is $600$ mJy Hz$^{-1/2}$. (This is the NEFD 
of the MPIfR 19 channel bolometer array on the Heinrich Hertz 
Telescope. Note that this value of NEFD includes atmospheric loss 
and time spent on the reference position.) To determine the
period of the flux variability, two independent measurements 
are needed over $P_{min}$. Setting the observed integration time
to $\tau = P_{min}/2$, an rms noise level of $25$~mJy can be
obtained, suggesting the predicted variability can be detected
at $\ge 3\sigma$ level in one 20 minute scan. Multiple 
scans can be co-added to improve the signal to noise ratio.
Ultimately, it may be necessary to use a mm/sub-mm interferometer
to resolve-out continuum emission not associated with Sgr A*. 

This sensitivity represents a fluctuation amplitude of $1-2\%$ in
Sgr A*'s average mm flux.  Previous monitoring of this source has
yielded only upper limits to the variability, though well above this
level.  In their multi-wavelength campaign, Falcke et al. (1998)
deduced a variability amplitude in Sgr A* of no larger than $20\%$
over several days at $3$ and $2$ mm.  Earlier, Gwinn et al. (1991)
searched for refractive scintillation arising from broadening in
the ISM, and inferred an upper limit of $\sim 10\%$ variability
at $0.8$ and $1.3$ mm on time scales of minutes to days.  The
present capability of bolometric systems on mm/sub-mm telescopes
appears to be considerably improved, and new observations should 
improve these limits by factors of five or better.

In this paper, we have focused primarily on the possibility of 
measuring Sgr A*'s light curve at mm/sub-mm wavelengths with sufficient
precision for us to infer the period associated with the marginally-stable 
orbit.  A power-density spectrum showing power at one or more QPO 
frequencies could in principle provide additional information on the gas 
dynamics near the black hole.  A detection of a periodic modulation in the 
light curve of Sgr A*, or of QPOs associated with its mm/sub-mm emission, 
would add considerably to our data base in a currently unexplored region of 
the spectrum.  Given that the {\it Chandra}-detected X-rays from this
source apparently scale strongly with the radio emission, periodic or
QPO signals may also be evident in the X-ray component.  Very importantly, 
these measurements would demonstrate convincingly that the radio emission 
from this source is associated with a nuclear Keplerian flow, whose 
gravitational confinement is provided by a supermassive compact object.  
And it may lead to the measurement of this object's spin with unprecedented 
accuracy. 

{\bf Acknowledgments} 
This research was partially supported
by NASA under grants NAG5-8239 and NAG5-9205, and has made use of NASA's
Astrophysics Data System Abstract Service.  FM is very grateful
to the University of Melbourne for its support (through a
Miegunyah Fellowship). 

{}

% And finally, we must deal with the figures.  There are three figures
% associated with this manuscript; two figures are Encapsulated
% PostScript (EPS) files.  The third figure is a grey scale figure that does
% not exist in EPS form.
%
% Authors have three options for including figure information within a 
% manuscript.  Not all the options may be acceptable by the target Journal - be
% sure to look at the appropriate submission instructions, electronic or 
% otherwise.
%
% Option 1.  Using this option, only the figure captions are included in the
% main body of the manuscript.  The figure captions must start on a new page.
% The captions are generated with the \figcaption[]{} command: the first 
% argument is optional, if you put something in there, put the name of the 
% EPS file that goes with the caption; the second argument is the figure 
% caption itself, and may include a \label command.  The \figcaption command
% generates the figure numbers.  This option is acceptable for all manuscript
% submissions.
%\newpage

\begin{figure}[thb]\label{fig1.ps}
{\begin{turn}{0}
\epsscale{0.8}
\centerline{\plotone{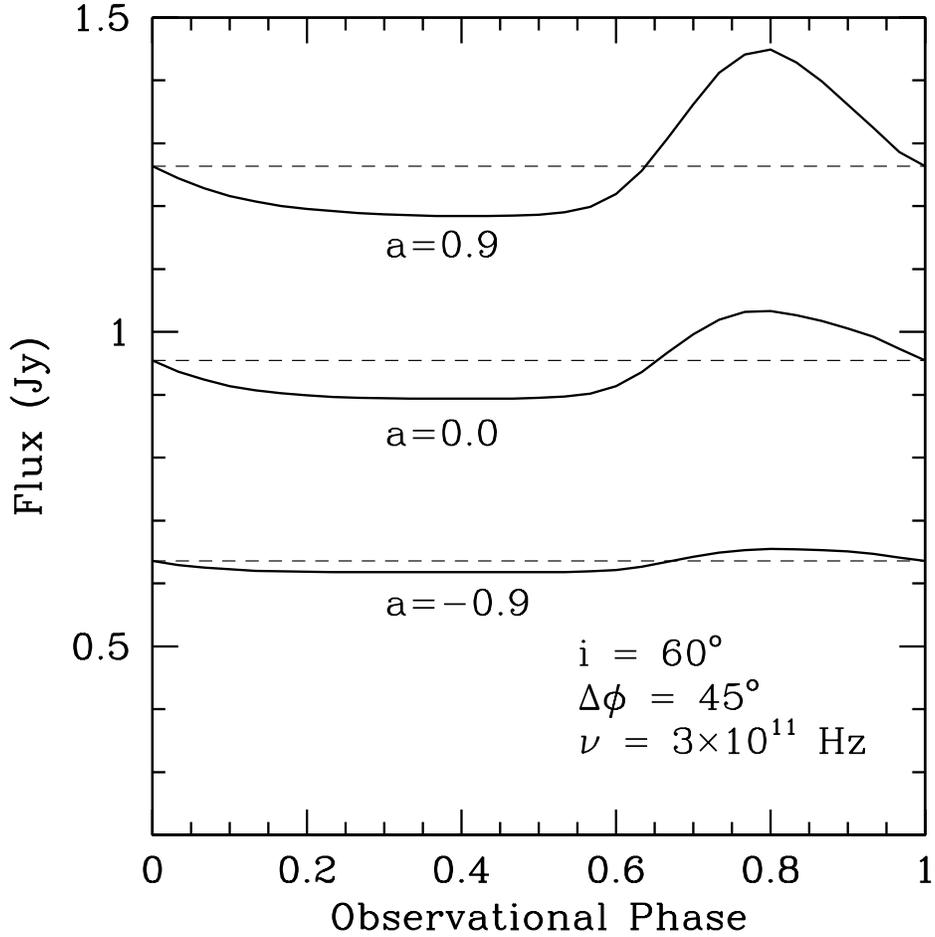}}
\end{turn}}
\caption{Flux density at $\nu=3\times 10^{11}$ Hz plotted as a function
of the observational phase for a $2.6\times 10^6\,M_\odot$ black hole
accreting at $4\times 10^{15}$ g s$^{-1}$, with a black hole spin $a=0$ 
(middle curve) and $a=0.9$ (top curve).  All curves are for a ``hot'' 
wedge ($\Delta\phi=45^\circ$), within which the temperature is artificially 
raised by $30\%$ above the value at other azimuthal angles. The assumed
inclination angle is $i=60^\circ$, and the outer disk radius is taken
to be $5\,r_S$, in accordance with the best fit model of Melia, Liu,
\& Coker (2000). The wedge extends from $r=3\,r_S$ to $5\,r_S$ in the
case of $a=0$, from $r=1.5\,r_S$ to $3.5\,r_S$ for $a=0.9$, and
from $r=4\,r_S$ to $5\,r_S$ when $a=-0.9$. For this particular simulation, 
the peak to trough variation in flux can be as high as $10\%$ over one 
complete cycle, though it tends to be smaller for the retrograde orbits.
Similar variations may be evident in the X-ray component of Sgr A*'s
spectrum if, as suggested by recent {\it Chandra} observations, the high-energy
photons are produced by self-Comptonization in the radio-emitting region.} 
\end{figure}


\begin{thebibliography}{}
\bibitem[Abramowicz et al. 1991]{A91}{Abramowicz, M.A., 
Bao, G., Lanza, A., \& Zhang, X.-H. 1991, AA, 245, 454}
\bibitem[Aitken et al. 2000]{Ait00}{Aitken, D.K., Greaves, J.,
Chrysostomou, A., Jenness, T., Holland, W., Hough, J.H., Pierce-Price, 
D. \& Richer, J. 2000, ApJ Letters, 534, L173}
\bibitem[Alexander \& Sternberg 1998] {AS98} Alexander, T.
\& Sternberg, A. 1998, ApJ, submitted.
\bibitem[Backer et al. 1993]{Bac93}{Backer, D.C., Zensus, J.A., 
Kellermann, K.I., Reid, et al. 1993, Science, 262, 1414}
\bibitem[Baganoff et al. 2001]{Bag01}{Baganoff, F. et al. 2001, ApJ,
submitted}
\bibitem[Bardeen, et al. 1972] {B72} Bardeen, J.M., Press, W.H.,
\& Teukolsky, S.A. 1972, ApJ, 178, 347.
\bibitem[Beckert \& Duschl 1997]{BD97}{Beckert, T. \& Duschl, W.J. 1997,
A\&A, 328, 95}
\bibitem[Biehle \& Blandford 1993]{bb93}{Biehle, G.T. \& Blandford, R.D. 1993, ApJ,
411, 302}
\bibitem[Bower et al. 1999]{Bow99}{Bower, G.C., Backer, D.C., Zhao, J.-H.,
Goss, M. \& Falcke, H. 1999, ApJ, 521, 582}
\bibitem[Bromley, Chen \& Miller 1997]{BroCheMil97}
{Bromley, B.C., Chen, K., \& Miller, W.A. 1997, ApJ, 475, 57}
\bibitem[Bromley, Melia \& Liu 2001]{BML01}{Bromley, B.C., Melia, F.,
and Liu, S. 2001, ApJ Letters, to be submitted}
\bibitem[Coker \& Melia 1997]{CM97}{Coker, R.F. \& Melia, F. 1997, ApJ Letters,
488, L149}
\bibitem[Coker \& Melia 2000]{CM00}{Coker, R.F. \& Melia, F. 2000,
ApJ, 534, 723}
\bibitem[Cui et al. 1998]{c98}{Cui, W., Zhang, S.N., \& Chen, W.
1998, ApJ, 492, L53}
\bibitem[Cunningham 1975]{Cun75}{Cummingham, C.T. 1975, ApJ, 202, 788}
\bibitem[Eckart, Genzel \& Hofmann 1995]{EGH95}{Eckart, A.,
Genzel, Hofmann, R., Sams, B.J. \& Tacconi-Garman, L.E. 1995, ApJ Letters,
445, L23}
\bibitem[Falcke et al. 1998]{Fal98}{Falcke, H., Goss, W.M., Matsuo, H.,
Teuben, P., Zhao, J.-H. \& Zylka, R. 1998, ApJ, 499, 731}
\bibitem[Genzel et al. 1996]{Gen96}{Genzel, R., Thatte, N., Krabbe, A.,
Kroker, H. \& Tacconi-Garman, L. E. 1996, ApJ, 472, 153}
\bibitem[Ghez et al. 1998]{Gh98}{Ghez, A.M., Klein, B.L., Morris, M. \&
Becklin, E.E. 1998, ApJ, 509, 678}
\bibitem[Gotthelf, Patterson \& Stover 1991]{GPS91}{Gotthelf, E., Patterson, J.,
\& Stover, R. J. 1991, ApJ, 374, 340}
\bibitem[Gwinn et al. 1991]{g91}{Gwinn, C.R., Danen, R.M., et al. 1991,
ApJ Letters, 381, L43}
\bibitem[Hawley et al. 1996] {HGB96} {Hawley, JF., Gammie, CF.,
Balbus, SA. 1996, ApJ, 464, 690}
\bibitem[Hollywood, et al. 1995] {HM95} Hollywood, J.M., Melia, F.,
Close, L.M., McCarthy, D.W., Jr., \& DeKeyser, T.A.
1995, ApJ Letters, 448, L21.
\bibitem[Kaaret et al. 1999]{k99}{Kaaret, P., Piraino, S., Bloser, P.F.,
Ford, E.C., Grindlay, J.E., et al. 1999, ApJ Letters, 520, L37}
\bibitem[Klu\'zniak \& Wagoner 1985]{k85}{Klu\'zniak, W. \& Wagoner, R.V. 1985, ApJ,
297, 548}
\bibitem[Kojima 1991]{Koj91}{Kojima, Y. 1991, MNRAS, 250, 629}
\bibitem[Kowalenko \& Melia 2000] {KM00} {Kowalenko, V, \& Melia, F.
2000, MNRAS, 310, 1053}
\bibitem[Krichbaum et al. 1993]{Kr93}{Krichbaum, T.P. et al. 1993, A\&A, 274, L37}
\bibitem[Krichbaum et al. 1998]{Kr98}{Krichbaum, T.P. et al. 1998, A\&A, 335, L106}
\bibitem[Lo et al. 1993]{Lo93}{Lo, K.Y., Backer, D.C., Kellermann, K.I., 
Reid, M., Zhao, J.H., Goss, W.M. \& Moran, J.M. 1993, Nature, 362, 38}
\bibitem[Madejski, et al. 1993] {M93} Madejski, G.M., et al.
\bibitem[Melia 1992]{Me92}{Melia, F. 1992, ApJ Letters, 387, L25}
\bibitem[Melia 1994]{Me94}{Melia, F. 1994, ApJ, 426, 577}
\bibitem[Melia, Jokipii \& Narayanan 1992]{MJN92}{Melia, F., Jokipii, J.R.
\& Narayanan, A. 1992, ApJ Letters, 395, L87}
\bibitem[Melia, Liu \& Coker 2000]{MLC00a}{Melia, F., Liu, S. \& Coker,
R.F. 2000, ApJ Letters, 545, L117}
\bibitem[Melia, Liu \& Coker 2001]{MLC00b}{Melia, F., Liu, S. \& Coker,
R.F. 2000b, ApJ, in press}
\bibitem[Melia \& Zylstra 1992]{MZ92}{Melia, F. \& Zylstra, G. 1992, ApJ Letters, 398, L53}
\bibitem[Menten et al. 1997]{Men97}{Menten, K.M., Reid, M.J., Eckart, A.
\& Genzel, R. 1997, ApJ Letters, 475, L111}
\bibitem[Miller et al. 1998]{m98}{Miller, M.C., Lamb, F.K., \& Psaltis, D. 
1998, ApJ, 508, 791}
\bibitem[Morgan et al. 1997]{m97}{Morgan, E.H., Remillard, R.A., \& Greiner, J. 
1997, ApJ, 482, 993}
\bibitem[Najarro, et al. 1997] {N97} Najarro, F., Krabbe, A.,
Genzel, R., Lutz, D., Kudritzki, R.P., \& Hillier, D.J.
1997, AA, 325, 700.
\bibitem[Narayan, Yi \& Mahadevan 1995]{NYM95}{Narayan, R., Yi, I. 
\& Mahadevan, R. 1995, Nature, 374, 623}
\bibitem[Nayakshin \& Melia 1997]{NM97}{Nayakshin, S. 
\& Melia, F. 1997, ApJ, 490, L13}
\bibitem[Nowak et al. 1997]{n97}{Nowak, M.A., Wagoner, R.V., Begelman, M.C., 
\& Lehr, D.E. 1997, ApJ Letter, 477, L91} 
\bibitem[Pacholczyk 1970]{Pac70}{Pacholczyk, A.G. 1970, Radio Astrophysics 
(W.H. Freeman and Company: San Francisco)}
\bibitem[Paczy\'nski 1987]{p87}{Paczy\'nski, B. 1987, Nature, 327, 303}
\bibitem[Papadakis \& Lawrence 1993] {PL93} Papadakis, I.E. \&
Lawrence, A. 1993, Nature, 361, 233.
\bibitem[Patterson 1981]{P81}{Patterson, J. 1981, ApJ Supplements, 45, 517}
\bibitem[Predehl \& Tr\"umper 1994]{PT94}{Predehl, P. \& Tr\"umper, J. 1994,
A\&A, 290, L29}
\bibitem[Stella \& Vietri 1998]{sv98}{Stella, L. \& Vietri, M. 1998,
ApJ Letters, 492, L59}
\bibitem[Strohmayer et al. 1996]{s96}{Strohmayer, T.E., Zhang, W., Swank, J.H., 
Smale, A., Titarchuk, L. Day, C. 1996, ApJ Letters, 469, L9}
\bibitem[van der Klis 1996]{v96}{van der Klis, M., Swank, J.H., Zhang, W., 
et al. 1996, ApJ Letters, 469, L1}
\bibitem[Zhang et al. 1997]{z97}{Zhang, W., Strohmayer, T.E., \& Swank, J.H. 1997,
ApJ Letters, 482, L167}
\bibitem[Zhang et al. 1998]{z98}{Zhang, W., Smale, A.P., Strohmayer, T.E.,
\& Swank, J.H. 1998, ApJ, 500, L171}
\end{thebibliography}
\end{document}